\documentclass[twocolumn,prl,superscriptaddress]{revtex4}   

\usepackage{dcolumn}
\usepackage{amsmath}

\usepackage{graphicx}

\newcommand{\dd}{\mathrm{d}}
\newcommand{\e}{\mathrm{e}}
\newcommand{\half}{\tfrac12}
\newcommand{\etal}{{\it{}et~al.}}
\newcommand{\cin}{c_\textrm{in}}
\newcommand{\cout}{c_\textrm{out}}

\begin{document}

\title{Estimating the number of communities in a network}

\author{M. E. J. Newman}
\affiliation{Department of Physics, University of Michigan, Ann Arbor, Michigan, USA}
\affiliation{Rudolph Peierls Centre for Theoretical Physics, University of Oxford, 1 Keble Rd., Oxford, UK}
\author{Gesine Reinert}
\affiliation{Department of Statistics, University of Oxford, 24--29 St.\,Giles, Oxford, UK}

\begin{abstract}  
Community detection, the division of a network into dense subnetworks with only sparse connections between them, has been a topic of vigorous study in recent years.  However, while there exist a range of powerful and flexible methods for dividing a network into a specified number of communities, it is an open question how to determine exactly how many communities one should use.  Here we describe a mathematically principled approach for finding the number of communities in a network using a maximum-likelihood method.  We demonstrate the approach on a range of real-world examples with known community structure, finding that it is able to determine the number of communities correctly in every case.
\end{abstract}

\maketitle

The large-scale structure of empirically observed networks, such as social, biological, and technological networks, is often complex and difficult to comprehend~\cite{Newman10}.  Community detection, the division of the nodes of a network into densely connected groups with only sparse between-group connections, is one of the most effective tools at our disposal for reducing this complexity to a level where network topology can be more easily understood and interpreted.  The development of algorithmic methods for community detection has been the subject of a large volume of recent research~\cite{GN02,Fortunato10,CGP11}, as a result of which we now have a number of efficient and sensitive detection techniques that are able to find meaningful communities in real-world settings~\cite{CGP11,Newman04a,PL05,RB07b,BGLL08,AK08,ABL10,KN11a}.

A fundamental limitation of most of these methods, however, is that they only divide networks into a fixed number of groups, so that one must know in advance how many groups one is looking for.  Normally one does not have this information, which significantly diminishes the usefulness of community detection as an analytic tool.  In this paper, we present a rigorous, first-principles solution to this problem in the form of an algorithm that, when applied to a given network, returns the number of communities the network contains.  The algorithm makes use of widely accepted methods of statistical inference coupled with a numerical approach that scales efficiently to large networks.

There have been a number of previous approaches proposed for this problem, among which perhaps the best known is the method of modularity maximization~\cite{NG04,Newman04a}, which is a method both for choosing the number of communities and for performing the community division itself.  This method is employed in, for example, the widely used Louvain algorithm~\cite{BGLL08}, but it suffers from being only heuristically motivated and there are instances where it is known to give incorrect results~\cite{FB07,GDC10}.  More rigorous approaches include the maximization of various approximations to integrated data likelihoods for generative network models, including Laplace-style approximations~\cite{DPR07}, variants of the Bayesian information criterion~\cite{HR07,LBA09}, and variational approximations~\cite{LBA12}.  Perhaps most similar to our work is that of~\cite{CL16} which uses an exact integral of the likelihood for a stochastic block model, as we do, but makes a number of other approximations and also employs a non-degree-corrected model, making it unsuitable for applications to most real-world network data.  Also of note is the minimum description length method of~\cite{Peixoto14a}, which at first sight is based on different ideas but can be shown to be equivalent to maximizing an integrated likelihood, though it uses a different model and different numerical methods~\cite{Peixoto15}.

Our approach, like much of the recent work in this area, is based on methods of statistical inference, in which one defines a model of a network with community structure, then fits that model to observed network data.  The parameters of the fit tell us about the community structure in much the same way that the fit of a straight line through a set of data points can tell us about their slope.  The model most commonly employed in this context is the stochastic block model~\cite{KN11a,HLL83,BC09}.  In this model one specifies the number of nodes~$n$ in the network along with the number~$k$ of communities or groups, then one assigns each node in turn to one of the groups at random, with probability $\gamma_r$ of assignment to community~$r$ (where $r$ runs from 1 to~$k$).  Note that we must have $\sum_{r=1}^k \gamma_r = 1$ for consistency.  Once all nodes have been assigned to a group, one places undirected edges independently at random between pairs of distinct nodes with probabilities~$\omega_{rs}$, where $r$ and $s$ are the groups to which the nodes belong.  If the diagonal parameters~$\omega_{rr}$ are greater than the off-diagonal ones, this produces a network with traditional community structure.

In practice, this model is often studied in a slightly different formulation in which one places not just a single edge between any pair of nodes~$i,j$ but a Poisson distributed number with mean~$\omega_{rs}$, or half that number when $i=j$~\cite{KN11a}.  In this variant of the model the generated network may contain both multiedges and self-edges, which is in a sense unrealistic---most real networks contain neither.  But in typical situations the edge probabilities are so small that both multiedges and self-edges occur with very low frequency, and the model is virtually identical to the first (Bernoulli) formulation given above.  At the same time the Poisson formulation is significantly easier to treat mathematically.  In this paper we use the Poisson version.

This definition of the model specifies its behavior in the ``forward'' direction, for the generation of random artificial networks, but our interest here is in its use in the reverse direction for inference, where we hypothesize that an observed network was generated using the model and then estimate by looking at the network which parameter values must have been used in the generation~\cite{NS01,BC09}.

Let the observed network be represented by its adjacency matrix~$A$, with elements~$a_{ij}=1$ if distinct nodes $i$ and $j$ are connected by an edge (or, by convention, $a_{ii}=2$ for self-edges) and $a_{ij}=0$ if nodes are not connected, and let the assignment of nodes to groups be represented by a vector~$g$ with elements $g_i$ equal to the group to which node~$i$ is assigned.  Then the probability, or likelihood, that the model generates a particular network~$A$ and group assignment~$g$, given the parameters $\gamma$, $\omega$, and~$k$, is
\begin{align}
P(A,g|\gamma,\omega,k) &= P(g|\gamma,k) P(A|g,\omega) \nonumber\\
  &\hspace{-6em}{} = \prod_i \gamma_{g_i} \prod_i \bigl( \half\omega_{g_ig_i}
     \bigr)^{a_{ii}/2} \e^{-\omega_{g_ig_i}/2}
     \prod_{i<j} \omega_{g_ig_j}^{a_{ij}} \e^{-\omega_{g_ig_j}} \nonumber\\
  &\hspace{-6em}{} = \prod_r \gamma_r^{n_r} \prod_r \omega_{rr}^{m_{rr}} \e^{-n_r^2\omega_{rr}/2}
  \prod_{r<s} \omega_{rs}^{m_{rs}} \e^{-n_r n_s\omega_{rs}},
\label{eq:likelihood}
\end{align}
where $n_r = \sum_i \delta_{g_i,r}$, is the number of nodes in group~$r$ (with $\delta_{ij}$ being the Kronecker delta), and $m_{rs}$ is the number of edges running between groups~$r$ and~$s$, given by $m_{rs} = \sum_{ij} a_{ij} \delta_{g_i,r} \delta_{g_j,s}$ for $r\ne s$ or half that number when~$r=s$.  (We have neglected an overall multiplicative constant in~\eqref{eq:likelihood}, since it cancels out of later calculations.)  Note that there is no requirement that all $k$ groups be non-empty: $k$~represents the number of groups nodes can potentially occupy, not the number they actually do.  Indeed it is crucial to allow for the possibility of empty groups for our calculations to be correct.

We can use Eq.~\eqref{eq:likelihood} to derive the probability $P(k,g|A)$ that, given an observed network~$A$, the block model from which it was generated had $k$ groups and group assignment~$g$, by an exact integral over the parameters~\cite{GS09,Peixoto14a,CL16}.  We assume maximum-entropy (least informative) prior probability distributions on the unknown quantities $k$, $\gamma$, and~$\omega$, which implies for instance that the prior on $k$ is uniform between the minimum and maximum allowed values of $k=1$ and $k=n$, meaning that $P(k)=1/n$, independent of~$k$.  The prior on the group assignment probabilities~$\gamma$ is also uniform, but because of the constraint $\sum_r \gamma_r = 1$ it occupies a more complicated space, a regular simplex with $k$ vertices and volume~$1/(k-1)!$, so that the prior probability density is $P(\gamma|k) = (k-1)!$.  For $\omega$ we set the scale of the prior (and hence the density of the network) by requiring that the mean of the edge probability~$\omega_{rs}$ be equal to the observed average edge probability in the network as a whole $p = 2m/n^2$, where $m$ is the observed number of edges in the network.  Then the maximum-entropy prior is an exponential $P(\omega) = p^{-1} \e^{-\omega/p}$.  (Approaches of this kind, where the prior is chosen to match features of the input data, are known as ``empirical Bayes'' techniques and typically give consistent results in the large-$n$ limit~\cite{CL08,PRS14}.)

Given the prior probabilities, we now have
\begin{equation}
P(k,g|A) = {P(k) P(g|k) P(A|g)\over P(A)},
\label{eq:kga}
\end{equation}
where
\begin{align}
\label{eq:pg}
P(g|k) &= \int P(g|\gamma,k) P(\gamma|k) \>\dd\gamma
        = {(k-1)!\over(n+k-1)!} \prod_{r=1}^k n_r! \\
\label{eq:pa}
P(A|g) &= \int P(A|g,\omega) P(\omega) \>\dd\omega \nonumber\\
       &\hspace{-2em}{} = \prod_r {m_{rr}!\over
          (\half p n_r^2+1)^{m_{rr}+1}}
          \prod_{r<s} {m_{rs}!\over(pn_r n_s+1)^{m_{rs}+1}}.
\end{align}
The probability~$P(A)$ in the denominator of~\eqref{eq:kga} is unknown but cancels out of later calculations (and we have again neglected an overall multiplicative constant in~\eqref{eq:pa}, for the same reason).

We can regard the values $k,g$ as defining a ``state'' of a statistical mechanical system with probability~$P(k,g|A)$.  We will sample states of this system in proportion to this probability using Markov chain Monte Carlo importance sampling~\cite{NB99,RC10}.  Then an estimate of the probability~$P(k|A)$ of having $k$ communities given the observed network~$A$ is given by the histogram of values of $k$ over the Monte Carlo sample, and the most likely value of~$k$ is the one for which $P(k|A)$ is greatest (although in many cases the complete distribution over $k$ can offer more insight than just its largest value alone).

This defines the method for estimating the number of groups~$k$.  It remains only to choose the Monte Carlo procedure.  In order to sample over both $k$ and $g$ we use two different Monte Carlo steps.

To sample over group assignments $g$ for given~$k$, we perform steps consisting of the movement of a single node from one group to another.  One could perform such steps using the classic Metropolis--Hastings rejection scheme, but we have found better efficiency (especially for larger values of~$k$) with a so-called heat-bath algorithm~\cite{NB99}, in which a randomly chosen node~$i$ is assigned a new group~$r$ from among the $k$ possibilities with probabilities $P(g_i=r|k,A) = P(k,g_i=r|A)/\sum_s P(k,g_i=s|A)$, all other $g_i$ being held constant.

To sample values of~$k$ with $g$ held constant we perform steps in which the value of $k$ is either increased or decreased by~1.  Using Eqs.~\eqref{eq:kga} and~\eqref{eq:pg}, the probabilities $P(k,g|A)$ and $P(k+1,g|A)$ are related by
\begin{align}
{P(k+1,g|A)\over P(k,g|A)}
  &= {P(k+1) P(g|k+1) P(A|g)/P(A)\over P(k) P(g|k) P(A|g)/P(A)} \nonumber\\
  &= {k!/(n+k)!\over(k-1)!/(n+k-1)!} = {k\over n+k},
\end{align}
where we have made use of the fact that $P(k)=1/n$ is independent of~$k$.  Thus, an appropriate Monte Carlo step is one in which with equal probability we propose either to decrease or increase~$k$ by~1; moves $k\to k-1$ are always accepted (provided they are possible at all, i.e.,~whenever $g$ has $k-1$ or fewer non-empty groups), and moves $k\to k+1$ are accepted with probability $k/(n+k)$.

This procedure constitutes a complete algorithm for determining the best-fit value of~$k$ but, helpful though it is as an illustration of the proposed method, it turns out to perform poorly in most real-world situations, for well-understood reasons.  The ordinary stochastic block model used here is known to give a poor fit, and hence poor results, for most real-world network data, because it fails to match the broad degree distributions commonly observed in such data~\cite{KN11a,CL16}.  The solution to this problem is to use a more elaborate model, the degree-corrected stochastic block model, which is able to fit networks with any degree distribution.  In this model one defines an additional set of continuous-valued node parameters~$\theta_i$, one for each node~$i$, and the expected number of edges between any pair of nodes $i,j$ becomes $\theta_i\theta_j\omega_{rs}$, where again $r$ and $s$ are the groups to which the nodes belong.  As discussed in~\cite{KN11a}, the parameters~$\theta_i$ allow us to independently control the average degree of each node and hence match any desired distribution, while the parameters~$\omega_{rs}$ control the community structure as before.

The model is not yet completely specified, however, because there is an arbitrary constant in the definition of~$\theta_i$: if we increase all the $\theta_i$ in group~$r$ by a factor of~$c_r$ and correspondingly decrease all $\omega_{rs}$ by a factor of~$c_rc_s$, the probability distribution over networks remains the same, regardless of the values of the~$c_r$.  In the language of statistics, the model parameters are not identifiable.  To fix the arbitrary constants one must specify a normalization for the~$\theta_i$ in each group, which can be done in a variety of ways.  In our work we impose the condition that the average value of $\theta_i$ be~1 in every group:
\begin{equation}
{1\over n_r} \sum_i \theta_i\,\delta_{g_i,r} = 1,
\label{eq:sumrule}
\end{equation}
for all~$r$.  This choice is convenient, since it has the effect of making the average number of edges between two different groups $r$ and~$s$ equal to $\sum_{ij} \theta_i \theta_j \omega_{rs}\,\delta_{g_i,r} \delta_{g_j,s} = n_r n_s \omega_{rs}$.  In other words, with this choice $\omega_{rs}$ represents the average probability of an edge between nodes in groups~$r$ and~$s$, just as it does in the standard stochastic block model.

With these definitions, the likelihood of a network~$A$ within the degree-corrected model, given a group assignment~$g$ and parameter sets~$\theta,\omega$, is
\begin{align}
P(A|g,\theta,\omega) &= \prod_i (\half \theta_i^2 \omega_{g_ig_i})^{a_{ii}/2}
  \e^{-\theta_i^2 \omega_{g_ig_i}/2} \nonumber\\
  &\hspace{4em}{}\times\prod_{i<j} (\theta_i\theta_j \omega_{g_ig_j})^{a_{ij}}
  \e^{-\theta_i\theta_j \omega_{g_ig_j}} \nonumber\\
  &\hspace{-4em}{} = \prod_i \theta_i^{d_i} \prod_r \omega_{rr}^{m_{rr}} \e^{-n_r^2\omega_{rr}/2}
  \prod_{r<s} \omega_{rs}^{m_{rs}} \e^{-n_r n_s\omega_{rs}},
\end{align}
where $d_i=\sum_j a_{ij}$ is the observed degree of node~$i$, we have used~\eqref{eq:sumrule} in the second equality, and we have again neglected an unimportant multiplicative constant.

We assume maximum-entropy priors as before, which again implies an exponential distribution $p^{-1} \e^{-\omega/p}$ for~$\omega$.  For $\theta$ it implies a uniform distribution over the regular simplex defined by Eq.~\eqref{eq:sumrule}.  Integrating over $\theta$ and~$\omega$, we find the value of $P(A|g)$ in the degree-corrected model to be the same as that for the uncorrected model, Eq.~\eqref{eq:pa}, except for an extra multiplicative factor of $\prod_{r:n_r\ne0} n_r^{\kappa_r} (n_r-1)!/(n_r+\kappa_r-1)!$ where $\kappa_r = \sum_i d_i \delta_{g_i,r}$ is the sum of the degrees of the nodes in group~$r$.  All other formulas remain the same as for the uncorrected model.  Modest though the change in $P(A|g)$ might seem, it produces a substantial difference in the behavior of the model, giving us a method that now works well on networks with any degree distribution.

Implementation of the complete method is straightforward.  At each time-step we perform either a group-update Monte Carlo step with probability $1-q$ or a $k$-update step with probability~$q$, where $q=1/(n+1)$, so that one $k$-update is performed on average for every $n$ group updates (one ``sweep'' of the system in the language of Monte Carlo simulation).  Run time per sweep is linear in~$n$, and we typically perform a few thousand sweeps in total, recording the value of~$k$ at regular intervals.  The calculations for the figures in this paper took seconds to minutes per network on a standard desktop computer, depending on network size.  The largest system we have studied comprised about $100\,000$ nodes and $800\,000$ edges and required an hour of running time for $10\,000$ Monte Carlo sweeps.  On some networks, particularly those with very weakly connected communities, the algorithm can get stuck in metastable states, in which case faster equilibration may be achieved by performing repeated runs on the same network with random initial conditions and using results from the run that achieves the highest average likelihood.  The computer code for our implementation of the method is available on the web~\cite{code}.

\begin{figure}
\begin{center}
\includegraphics[width=\columnwidth,clip=true]{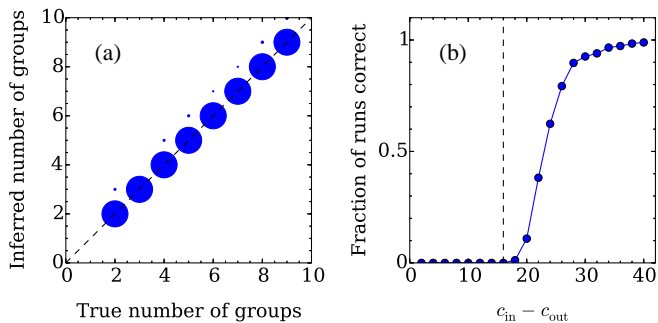}
\end{center}
\caption{Tests of the method on synthetic networks generated using the stochastic block model.  (a)~Diameter of points represents the likelihood $P(k|A)$ of inferred values of $k$ as a function of true~$k$ for networks with $k$ groups of size 250 nodes each.  Each node has an average of 16 edges connecting it to its own group and 8 edges to each other group.  For each value of~$k$ we performed 10 runs of 2000 Monte Carlo sweeps each (plus 1000 for equilibration) and took our results from the run that found the highest average likelihood.  Correct inference would place most weight along the dashed diagonal line.  (b)~The fraction of runs detecting the correct number of groups in stochastic block models with $k=4$ groups of 250 nodes each and average degree 16, as a function of the strength of the community structure.  The vertical dashed line represents the theoretical detectability threshold below which every algorithm must fail.  Each point is an average over 1000 networks and success is defined as assigning an absolute majority of the probability~$P(k|A)$ to the correct value of~$k$.}
\label{fig:synthetic}
\end{figure}

We have tested the method on a range of different networks, including computer-generated (``synthetic'') networks with known community structure as well as real-world examples.  Figure~\ref{fig:synthetic} shows results for synthetic networks generated using the standard (non-degree-corrected) stochastic block model with edge probabilities~$\omega_{rs}$ equal to $\cin/n$ when $r=s$ (in-group connections), $\cout/n$ when $r\ne s$ (between-group connections), and $\cin>\cout$, so that the network shows traditional assortative structure.  Figure~\ref{fig:synthetic}a shows results for the likelihood~$P(k|A)$ for networks with a range of values of $k$ and, as the figure shows, the algorithm overwhelmingly assigns highest likelihood to the correct value of~$k$ in every case.  We can make the problem more challenging by decreasing the difference $\cin-\cout$ between the numbers of in- and out-group connections, thereby generating networks with weaker community structure that should be harder to detect.  Typical community detection algorithms show progressively poorer performance as structure weakens and it can be proved that when it is sufficiently weak the structure becomes undetectable by any means, a phenomenon known as the detectability transition~\cite{DKMZ11a,MNS15}.  We see similar behavior in detecting the number of communities, as shown in Fig.~\ref{fig:synthetic}b, where we apply our algorithm to 1000 networks for each of several values of $\cin-\cout$ while holding $k$ fixed and plot the fraction of runs on which we arrive at the correct answer for the number of groups.  Below the detectability threshold the algorithm fails to determine the correct result, as all algorithms must, but as we move above the threshold performance improves and for larger values of $\cin-\cout$ the algorithm once again returns the correct answer on almost every run.

\begin{figure}
\begin{center}
\includegraphics[width=\columnwidth]{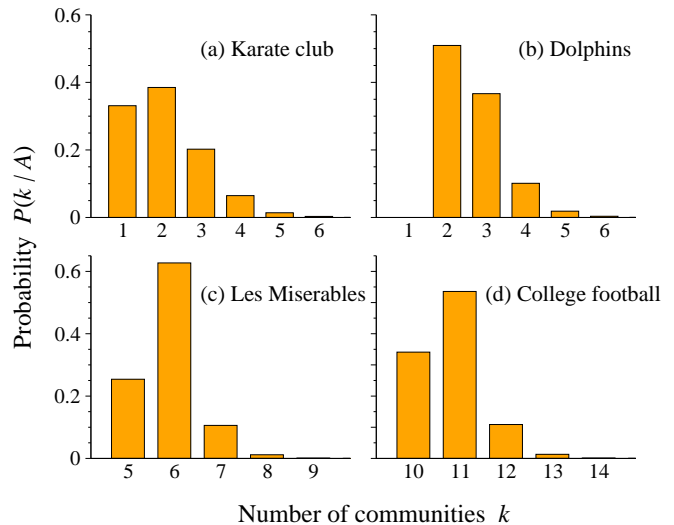}
\end{center}
\caption{Posterior probabilities~$P(k|A)$ calculated using the method of this paper for four real-world networks with known community structure, as described in the text.  For each network we performed 10 runs of $50\,000$ Monte Carlo sweeps each (plus $50\,000$ for equilibration), taking our results from the run that finds the highest average likelihood.}
\label{fig:real}
\end{figure}

Figure~\ref{fig:real} shows the results of tests of the algorithm on four real-world networks whose community structure is widely agreed upon: the well-studied ``karate club'' network of Zachary~\cite{Zachary77}, which is generally thought to have two groups; the dolphin social network of Lusseau~\etal~\cite{Lusseau03a}, also thought to have two groups; the co-appearance network of fictional characters in the novel \textit{Les Miserables} by Victor Hugo~\cite{NG04}, with six groups corresponding to major subplots of the story; and the network of games between Division I-A American college football teams in the year 2000~\cite{NG04}, with 11 groups corresponding to the established conferences of US collegiate sports competition (or, arguably, 12~if one includes the independent teams that do not belong to any conference).  The figure shows histograms of the estimated probabilities~$P(k|A)$ for each of these four networks and the peak probability falls at the agreed-upon value in each case---at $k=2$, 2, 6, and~11 respectively.  In each case the accepted value easily outweighs any other and the choice of group number is clear, except in the case of the karate club network, for which $k=2$ does receive the most weight but $k=1$ comes a close second.  This is an interesting finding in the context of this particular network, which comes from a study of a university student club that was a single group at the time the network was observed but broke into two shortly afterwards.  Our results fit this observation neatly, indicating that the network could be construed either as a single community or as a pair of communities.

Once the value of $k$ for a network has been determined, one does not necessarily need to perform a separate calculation to determine the community structure itself.  Since our Monte Carlo procedure samples group assignments~$g$ from the distribution~$P(k,g|A)$, one can simply examine the subset of sampled assignments corresponding to the inferred value of~$k$ to get an estimate of the posterior distribution over network divisions.  In particular, one can calculate the marginal probability that a node belongs to any given group to within an overall constant from $P(g_i=r|k,A) \propto \sum_g \delta_{g_i,r} P(k,g|A)$ and then assign each node to the group for which this probability is largest, obviating the need for other methods of fitting the block model, such as maximization of the profile likelihood~\cite{BC09,KN11a}.

In summary, we have given a first-principles method for inferring the number of communities into which a network divides.  In tests, the method, based on simultaneous Monte Carlo sampling of the distribution of community divisions and community number, gives correct answers on a range of benchmark networks with known community structure.  The method can be scaled up, without significant modification, to allow the analysis of data sets with hundreds of thousands of nodes or more.

\begin{acknowledgments}
The authors thank Tiago Peixoto and Maria Riolo for helpful conversations, and several anonymous referees for providing substantial and useful feedback.  This work was funded in part by the US National Science Foundation under grants DMS--1107796 and DMS--1407207 (MEJN), the UK Engineering and Physical Sciences Research Council under grant EP/K032402/1 (GR), and the Advanced Studies Centre at Keble College, Oxford.
\end{acknowledgments}

\end{document}